\documentclass[twocolumn, showpacs,preprintnumbers,amsmath,amssymb]{revtex4}

\usepackage{bm}

           \usepackage{graphicx}
           \usepackage{bm}
           \usepackage{epsfig}

\begin{document}

\title{Topological Disorder in Spin Models on Hierarchical Lattices.}

\author{N. S. Ananikian, K. G.~Sargsyan\\[1mm]
{\small \textsl{Department of Theoretical Physics, Yerevan Physics
Institute,}}\\{\small Alikhanian Br. 2, 375036, Yerevan, Armenia
}\\[1mm]}

\begin{abstract}
A general approach for the description of spin systems on
hierarchial lattices with coordination number $q$ as a dynamical
variable is proposed. The ferromagnetic Ising model on the Bethe
lattice was studied as a simple example demonstrating our method.
The annealed and partly annealed versions of disorder concerned with
the lattice coordination number are invented and discussed.
Recurrent relations are obtained for the evaluation of
magnetization. The magnetization is calculated for the particular
disorder choices, $q=2,3$ and $q=3,4$. A nontrivial localization of
critical point is revealed.
\end{abstract}
 \maketitle
\section{Introduction}
The hierarchical models in statistical physics can be used to
exactly describe the critical phenomena. For these purposes only
approximate or numerical methods exist in most statistical
mechanical models. Therefore complicated models representing
nontrivial critical behavior and demonstrating analytical solution
may be recognized as very important. A good example of such systems
are spin models on hierarchical lattices. Besides the general
interest related to the critical phenomena, spin models on
hierarchical lattices are also formulated for description of the
concrete physical systems. The interesting applications are Bethe
approximation in polymer physics, and many others
\cite{Pretti}-\cite{esk}.

 The main purpose of this work is a study of spin models on hierarchical lattices
with so called topological disorder. The problem of disorder for the
hierarchical models is known and has been discussed in series of
works \cite{dis1}-\cite{dis4}. However, disorders concerned with
couplings and magnetic fields are considered mainly. So called
topological disorder, describing by the changing lattice
coordination number, is investigated less. In our definition
topological disorder represents non-homogeneity of the lattice
structure and may be even non-connected to the spins dynamics. A
more precise definition of the topological disorder in context of
the hierarchical models is given below within a concrete example.

Although all results in our article are related to the physics and
have direct meaning in a physical sense, it is interesting to
discuss their possible applications in the context of social
dynamics. As is known, in the recent decades there is a rapidly
growing interest to the application of the statistical physics to
interdisciplinary fields (as biology, computer science, social
dynamics and e.t.c). Considering the social dynamics we deal not
with physical particles but with humans or crowds. The human
behavior is not rather well understood and depends on many hidden
factors. Nevertheless, one may suppose that in the statistical limit
the new statistical laws would be obtained, which are less dependent
on the individuals. As it was shown in recent decade, this
conjecture is correct. The transitions from disorder to order are
discovered describing the social dynamics from the statistical
physics point of view. We may bring as the examples of those
transitions spontaneous formation of a common language, culture or
the emergence of consensus about a specific issue \cite{soc}. There
are also examples of scaling and universality.

As an illustration of our approach to spin systems on hierarchical
lattices with changing coordination number $q$ we consider
ferromagnetic Ising model on the Bethe lattice as a simplest
possible example.

The article is organized as follows. In section \ref{second} a
simple ferromagnetic Ising model with topological disorder on
hierarchical lattice is formulated. A more precise and clear
definition of the topological disorder in context of the
hierarchical lattices is given. The connection between proposed
model and description of social dynamics is discussed in detail.

In section \ref{3} annealed and partly annealed topological
disorders are defined. A general method for those mathematical
description is proposed. The annealed version of the topological
disorder is studied for the particular realizations of the lattice
structure. The critical temperature is found and analyzed.

Partly annealed and quenched topological disorders are described
within ferromagnetic Ising model in section \ref{4}. The same
particular realizations of the topological disorder as for the
annealed one are discussed. The critical point is obtained.

Main observations are presented in conclusion.
\section{Ising Model on Bethe Lattice with topological
disorder}\label{second}
 The one of the simplest recursive lattices is a Cayley tree. Constructing Cayley
tree one starts with a central site \emph{O}. After that we connect
the central site \emph{O} by links with $q$ new other sites. This
procedure of $q$ new sites connection by links to the each site
originated on the previous step repeats $n$ times. Following this
procedure one gets the recursive connected graph, containing no
cycles and called the Cayley tree with coordination number $q$ and
$n$ generations.

The Bethe lattice may be considered as the interior of the Cayley
tree. Only sites lying deep inside Cayley tree constitute the Bethe
lattice. Dealing with the Bethe lattice we suppose the respective
Cayley tree is large enough to achieve thermodynamical limit and
neglect boundary effects. The general structure and topology of the
Bethe lattice is the same as for Calyley tree. Here we consider
changing along the lattice coordination number $q_{i}$, which is
different in general for the different $i$ sites, even from the same
generation. The Betthe lattice is not uniform now and topological
disorder related to it can be discussed.

 The Hamiltonian of the Ising model on Bethe lattice reads
\begin{equation}
H=-J\sum_{i,j}s_{i}s_{j}-h\sum_{i}s_{i},\label{hamilton}
\end{equation}
where $J$ and $h$ do not depend on the indexes $i,j$, for the sake
of simplicity. The summation is over all neighbor $s_{i}$ and
$s_{j}$ spins along the lattice and spin variables are $s_{i}=\pm1$.
The interpretation of this model in context of social dynamics is
following. The common task of social dynamics is the understanding
of the transition from an initial disordered state to a
configuration that displays order. The one approach is an analogy to
physics of the Ising model for ferromagnetic. The Ising model may be
considered as the simplest description of the opinion dynamics. Each
spin on the Bethe lattice represents a human opinion on some subject
(agree/disagree units)locating in corresponding lattice site. We
describe the situation in the simplest sense. The coupling $J$ is
positive reflecting the fact that the people communicating with each
other long or even not a very long time usually tend to have the
same opinion. In other words, humans or other social units like
communicate with people (social units) having usually a good
agreement with them. However, it is not always true. The discussion
of the anti-ferromagnetic $J$ is also meaningful. The Bethe lattice
in our interpretation is not a spatial distribution of humans but a
construction in relation space and represents possible "stable"
communications (relations) between "humans" $s_{i}$. Our proposition
is to consider "humans" $s_{i}$ communicating with (acting on) some
finite number of neighbors in relation space. This number, which
coincides with the coordination number of lattice $q_{i}$, may be
influenced by some external noise and changes in time in general.
Here we consider $q_{i}$ individually for the each site $i$ and the
coordination number is not the same for the all sites of the same
step of the lattice generation.

An external magnetic field may be treated as an external
informational factor, like TV or something similar. As we use
methods of the statistical physics, the temperature is introduced
and represents the noise in human opinion ( measure of uncertainty
in taking stable decision ). The main weakness of the model is an
absence of the small cycles (loops) in relation space. Some effects
induced by relation loops may be demonstrated on more complicated
hierarchical lattices (as Husimi lattice).
\section{Annealed topological disorder}\label{3}
 First of all let us recall some general aspects related to the spin systems with disorder.
Let the Hamiltonian may be written as $H[J,s]$, where $J$ represents
disorder degree of freedom and $s$ describes the spin variables
correspondingly. Suppose that the spin and disorder degrees of
freedom are not thermally equilibrated and disorder degrees of
freedom have their own temperature $\overline{T}$.  The partition
function reads:
\begin{eqnarray}
&\textbf{Z}&=\int DJ P(J) \exp(-\overline{\beta}F(J))\nonumber
\\&=&\int DJ P(J)\exp(\frac{\overline{\beta}}{\beta}\ln
Z(J))\nonumber \\&=&\int DJ P(J)(Z(J))^{n},\label{part}
\end{eqnarray}
where $Z(J)$ and $F(J)$ are respectively partition function and free
energy with given disorder $J$. Here we propose to describe the
disorder by the probability distribution $P[J]$. We may consider the
number $n=\frac{\overline{T}}{T}$ as a number of replicas within the
replica formalism \cite{Dotsenko}. The total free energy would be
\begin{equation}
\textbf{F}=-\frac{T}{n}\ln(<Z(J)^{n}>_{dis}), \label{total}
\end{equation}
where the average over. Using $\textbf{F}$ one would obtain the free
energy describing quenched disorder after taking the limit
$n\rightarrow0$. The $n=1$ value corresponds to annealed disorder,
when both temperatures $T$, $\overline{T}$ are equal. The value
$1>n>0$ may be treated here as a "degree of quenchness". The same
heuristics arises from the non-equilibrium two temperature system
viewpoint in \cite{Allahv}.

Now let us consider topological disorder given by the set of
$\{q_{i}\}$ variables. We describe only the case of slow changing
$q_{i}$ compared with the rate of the spin flips. This assumption is
in agreement with the ferromagnetic hypothesis of social opinion
dynamics.  It is natural to suppose that the stable relations
dispose to agreement between neighbors to a greater extent. So we do
not analyze a more fast rate of $q_{i}$ than it is for the annealed
disorder. Also we propose the ability of the probability
factorization $P(\{ q_{0},...,q_{M}\})=P(q_{0})...P(q_{M})$.

The one of the best methods developed for the models on hierarchical
lattices is a so-called dynamical approach
\cite{din0}-\cite{Monroe}. Here we propose to modify this formalism
for our purposes. The main quantity to be calculated is
$<Z(\{q_{i}\})^{n}>_{dis}$. Let us start from the description of the
annealed topological disorder $n=1$. The main advantage of the
dynamical approach is that for the models formulated on recursive
trees the exact recursion relations can be derived. When the tree is
cut apart at the central site $O$, it separates into $q_{0}$
identical branches. According to it for the Hamiltonian
(\ref{hamilton}) one can write
\begin{eqnarray}
\textbf{Z}=<Z(\{q_{i}\})>_{dis}=\sum_{q_{0}}P(q_{0})\sum_{s_{0}}\exp(\beta
h s_{0})\times \nonumber
\\<g_{k}^{(1)}(q_{0},s_{0})...g_{k}^{(q_{0})}(q_{0},s_{0})>_{dis},\label{dyn}
\end{eqnarray}
where independent replicas $g_{k}^{(i)}(q_{0},s_{0})$ are invented.
The Eq. (\ref{dyn}) also may be rewritten as
\begin{eqnarray}
<Z(\{q_{i}\})>_{dis}=\sum_{q_{0}}P(q_{0})\sum_{s_{0}}\exp(\beta
h s_{0})\times \nonumber
\\<g_{k}(s_{0})>_{dis}^{q_{0}},\label{dyn1}
\end{eqnarray}
where
\begin{eqnarray}
<g_{k}(s_{0})>_{dis}=\sum_{q_{0}}P(q_{0})\sum_{s_{1}}\exp(\beta J
s_{0}s_{1}+\beta h s_{1})\times \nonumber
\\<g_{k-1}(s_{1})>^{q_{1}-1}_{dis}.
\end{eqnarray}
After denoting $x_{k}=<g_{k}(+)>$ and $y_{k}=<g_{k}(-)>$ we obtain
nonlinear two dimensional mapping
\begin{eqnarray}
x_{k}=\sum_{q} P(q)\left[e^{\beta J+\beta h}x^{q-1}_{k-1}+e^{-\beta
J-\beta h}y^{q-1}_{k-1}\right]\nonumber\\
y_{k}=\sum_{q} P(q)\left[e^{-\beta J+\beta h}x^{q-1}_{k-1}+e^{\beta
J-\beta h}y^{q-1}_{k-1}\right],\label{mapp}
\end{eqnarray}
where $x_{k}$ and $y_{k}$ must be positive.

\begin{figure}
          \begin{center}
          \includegraphics[width=7.5cm]{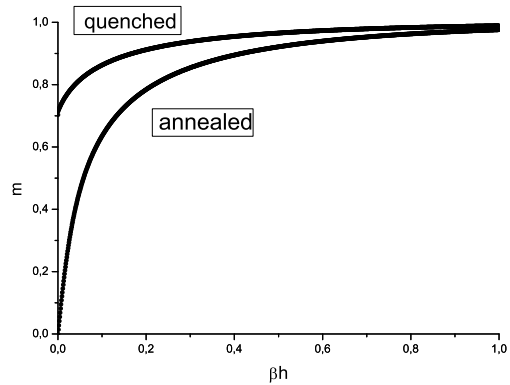}
          \caption{\label{magni} Magnetization curves with $\beta J=0.5$ in presence of topological disorder, where
          $q_{k}=3,4$ with equal probabilities.}
          \end{center}
        \end{figure}

As a possible order parameter one may consider magnetization
averaged over the disorder. This quantity represents averaged
opinion in terms of agree/desagree units. Using Eq. (\ref{part}) we
get magnetization of the central cite \emph{O} averaged over
topological disorder as
\begin{equation}
m_{dis}=\frac{\sum_{q_{0}}P(q_{0})\sum_{s_{0}}s_{0}e^{\beta h
s_{0}}<g_{k}(s_{0})>_{dis}^{q_{0}}}{\sum_{q_{0}}P(q_{0})\sum_{s_{0}}e^{\beta
h s_{0}}<g_{k}(s_{0})>_{dis}^{q_{0}}}.\label{magn}
\end{equation}
Getting fixed points $(\overline{x},\overline{y})$ of mapping
(\ref{mapp}) and following to Eq. (\ref{magn}) one would obtain the
order parameter $m_{dis}$
\begin{equation}
m_{dis}=\frac{\sum_{q_{0}}P(q_{0})\left[\exp(\beta h
)\overline{x}^{q_{0}}-\exp(-\beta h
)\overline{y}^{q_{0}}\right]}{\sum_{q_{0}}P(q_{0})\left[\exp(\beta h
)\overline{x}^{q_{0}}+\exp(-\beta h
)\overline{y}^{q_{0}}\right]}.\label{magn1}
\end{equation}
For the more concreteness we regarded the Bethe lattice where
$q_{i}$ takes values $q=3,4$ with equal probability on the each step
of the lattice generation and for the each site. Corresponding
magnetization $m_{dis}$ with respect to external field $\beta h$ at
$\beta J=0.5$ point is given in Fig. (\ref{magni}).
To obtain the magnetization we use the following numerical
procedure. At each point $\beta h$, starting from $\beta h=0$ with
step $\Delta(\beta h)=0.001$, we get fixed points of the mapping
(\ref{mapp}). In our simple example, evaluating fixed points one
have to replace ${\{x_{k},y_{k}\}}$ and ${\{x_{k-1},y_{k-1}\}}$ in
equations (\ref{mapp}) on ${\{x,y\}}$. Solution of the obtained
system of equations with respect to ${\{x,y\}}$ gets the set of the
searched fixed points. In presence of the magnetic field there are
only one fixed point holding conditions $x_{k}>0$ and $y_{k}>0$. In
case of the zero magnetic field one have two stable fixed points and
one nonstable, or otherwise, depending on temperature. For checking
stability of the fixed points of (\ref{mapp}), one can calculate
Jacobian of (\ref{mapp}) as
\begin{equation}
J=\left|
\begin{array}{ll}
 e^{J+h}x+\frac{3}{2}e^{J+h}x^2 & e^{-J-h}y+\frac{3}{2}e^{-J-h}y^2 \\
  e^{-J+h}x+\frac{3}{2}e^{-J+h}x^2 & e^{J-h}y+\frac{3}{2}e^{J-h}y^2
\end{array}
\right|. \label{jakob}
\end{equation}
The condition of stability now sounds as $|J|<1$.

It is known that the Ising model on Bethe lattice with constant $q$
has a critical point defined by condition $\beta_{c}
J=\frac{1}{2}\ln \frac{q}{q-2}$ \cite{Baxter}. According to
\cite{Baxter}, it is possible to write one dimensional mapping for
the partition function of the model with fixed $q$. This mapping has
three fixed points in zero magnetic field. It is easy to check that
the temperature corresponding to the case of the three non stable
fixed points defines by the same condition as the critical
temperature. When temperature is less then the critical temperature,
one obtains two stable fixed points (spontaneous magnetization) and
one non stable $x=1$. For the temperature greater then critical one,
we get two nonstable and one stable $x=1$ points. This
interpretation of known regimes \cite{Baxter} around the critical
point gives a possibility for the numerical investigation in
presence of the topological disorder. For the evaluation of critical
temperature of our example $q=3,4$ we propose the following
numerical procedure. Starting from point $\beta J=0$ and holding
$h=0$, with step $\Delta (\beta J)=0.005$, we check stability of
obtained fixed points and looking for $\beta J$ at which two stable
points become unstable and the third one becomes stable. The
obtained result is not an exact value of the critical temperature,
but an interval of possible values. The size of this interval
depends on the precision $\Delta (\beta J)$. As a result we found
$\beta_{c} J=0.725$, which is greater then the corresponding values
for the Ising models on Bethe lattice with fixed $q=3$ and $q=4$.

 Another interesting choice of topological disorder is $q=2,3$, where $q$
takes corresponding values with probability $P(q)$. Here we consider
only uniform probability distribution $P(2)=P(3)=0.5$. The general
study is a subject of further research. The coordination number
$q=2$ corresponds to the one dimensional Ising model having zero
critical temperature. The existence and localization of the critical
point in this mix of one-dimensional Ising model and Ising model on
Bethe lattice is a subject of special interest. The main idea is an
evaluation of the fixed points of corresponding mapping, as it was
done above for $q=2,3$. However, during this procedure one obtains
no fixed points holding condition $x>0,y>0$. For any positive
initial seed of mapping, after some finite number of iterations
(after $15-20$ iterations) $x_{k},y_{k}$ become infinitely big. To
deal with these big numbers we define new variable
$z_{k}=x_{k}/y_{k}$. The mapping (\ref{mapp}) may be represented now
as
\begin{equation}
z_{k}=\frac{P(3)(e^{\beta J+\beta h}z^{2}_{k-1}+e^{-\beta J-\beta
h})+o(n)}{P(3)(e^{-\beta J+\beta h}z^{2}_{k-1}+e^{\beta J-\beta
h})+o(n)}.\label{mappa}
\end{equation}
The infinitesimals $x_{k}/y^{2}_{k},1/y_{k}$ may be neglected and we
obtain that the critical temperature is the same as for Ising model
on Bethe lattice with constant $q=3$.
\section{Quenched topological
disorder}\label{4} In case of partly annealed topological disorder
$0<n<1$ we use the following heuristics. $\textbf{Z}$ quantity may
be represented as
\begin{eqnarray}
\textbf{Z}=<Z(\{q_{i}\})^n>_{dis}=\sum_{\{q_{i}\}}P(\{q_{i}\})[\exp(\beta
h )g_{\{q_{i}\}}^{q_{0}}(+)\nonumber\\+\exp(-\beta h
)g_{\{q_{i}\}}^{q_{0}}(-)]^n,\label{dyn2}
\end{eqnarray}
where $g_{\{q_{i}\}}(s_{0})$ depends on the particular realization
of disorder. One may obtain recursion relations from the previous
expression as
\begin{eqnarray}
x_{k}=e^{\beta J+\beta h}x^{q_{k}-1}_{k-1}+e^{-\beta
J-\beta h}y^{q_{k}-1}_{k-1}\nonumber\\
y_{k}=e^{-\beta J+\beta h}x^{q_{k}-1}_{k-1}+e^{\beta
J-\beta h}y^{q_{k}-1}_{k-1},\label{mapp2}
\end{eqnarray}
denoting $x_{k}=g_{\{q_{i}\}}(+)$ and $y_{k}=g_{\{q_{i}\}}(-)$. It
is assumed that on the each step of recursion the power $q_{k}$
takes random values with given probability distribution $P(q)$. The
main idea is to consider that although the mapping is stochastic in
nature, after some huge number of iterations one would obtain an
attractor set of possible pairs $(\overline{x},\overline{y})$ with
corresponding stationary probability distribution $P(x,y)$. To find
the attractor numerically one may study the mappings (\ref{mapp2})
with fixed non stochastic values of $q_{k}=q$. The fixed points of
those mappings should be a part of searching attractor. One may
start stochastic iteration (\ref{mapp2}) from those fixed points and
get the set of possible pairs $(\overline{x},\overline{y})$. There
is a possibility of a disconnected attractor (and broken
ergodicity), therefore the final set should be based on the
summation of sets derived from iterations of all possible fixed
points. During the iteration process one gets not only an attractor
set but also it is possible to evaluate $\textbf{Z}$. The method of
iterations takes a chance to calculate value of $<[\exp(\beta h
)x_{k}^{q_{0}}+\exp(-\beta h)y_{k}^{q_{0}}]^n>_{dis}$. Although the
numerical and approximate nature of such evaluation we suppose a
consistent result in the limit of large number of iterations. The
magnetization in presence of partly annealed disorder may be written
as
\begin{widetext}
\begin{eqnarray}
m_{dis}&=&\frac{\int DJ P(J)m(J)\exp(-\overline{\beta}F(J))}{\int DJ
P(J)\exp(-\overline{\beta}F(J))}=
\frac{\int DJ P(J)\frac{\partial Z(J)}{\partial \beta h}(Z(J))^{n-1}}{\int DJ P(J)(Z(J))^{n}}\nonumber\\
&=&\frac{<[\exp(\beta h )x_{k}^{q_{0}}-\exp(-\beta
h)y_{k}^{q_{0}}]\times[\exp(\beta h )x_{k}^{q_{0}}+\exp(-\beta
h)y_{k}^{q_{0}}]^{(n-1)}>_{dis}}{<\exp(\beta h
)x_{k}^{q_{0}}+\exp(-\beta h)y_{k}^{q_{0}}]^n>_{dis}}.\label{magn3}
\end{eqnarray}
\end{widetext}
However, the analysis of mapping (\ref{mapp2}) for the simple
example where $q_{k}=3,4$ with equal probabilities demonstrates some
difficulties. After few steps of iteration the values of
$x_{k},y_{k}$ approach to zero or infinity. To overcome it we
propose a mathematical trick, which should give correct values of
magnetization for $n\rightarrow0$ and exact magnetization in
presence of quenched topological disorder $n=0$. Instead of $x_{k}$
and $y_{k}$ one may consider new variable $z_{k}=x_{k}/y_{k}$ and
write a one dimensional mapping
\begin{equation}
z_{k}=\frac{e^{\beta J+\beta h}z^{q_{k}-1}_{k-1}+e^{-\beta J-\beta
h}}{e^{-\beta J+\beta h}z^{q_{k}-1}_{k-1}+e^{\beta J-\beta
h}}.\label{onemap}
\end{equation}
The mathematical trick lies in taking out the means over the
disorder $y_{k}$ in Eq. (\ref{magn3}), like it is possible to do
with constant parameter (real number). It is supposed that if
$y_{k}$ is infinite quantity or infinitesimal one, than this
approach is reliable. The magnetization may be rewritten now as
\begin{widetext}
\begin{equation}
m_{dis}\approx \frac{<[\exp(\beta h )z_{k}^{q_{0}}-\exp(-\beta
h)]\times[\exp(\beta h )z_{k}^{q_{0}}+\exp(-\beta
h)]^{(n-1)}>_{dis}}{<\exp(\beta h )z_{k}^{q_{0}}+\exp(-\beta
h)]^n>_{dis}}.\label{mag4}
\end{equation}
\end{widetext}
Previous relation Eq. (\ref{mag4}) becomes exact in presence of
quenched topological disorder as it follows directly from Eq.
({\ref{magn3}}). Using this trick is easy to get magnetization for
our example with $q=3,4$. To obtain magnetization we change the
values $\beta h$ like it was done for the annealed version of
disorder. We start evaluation of important means after $10^4$-th
iteration and calculating them using the same number of iterations.
The result does not change if this value is greater, therefore we
suppose that it is stable. The values of $z_{k}$ are distributed
around the fixed points of the mappings (\ref{onemap}) with fixed
values of $q_{k}$. The dependence of magnetization on external field
$h$ in presence of quenched topological disorder is given in Fig
(\ref{magni}).
\begin{figure}
          \begin{center}
          \includegraphics[width=7.5cm]{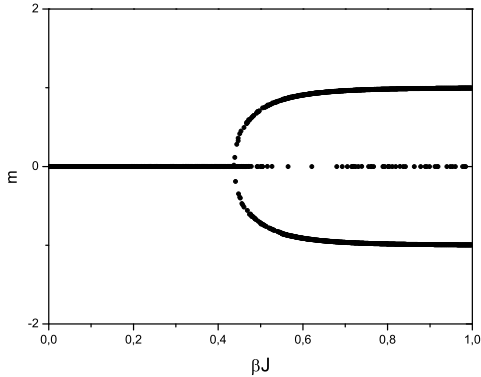}
          \caption{\label{crit} Spontaneous magnetization in presence of quenched topological disorder, where
          $q_{k}=3,4$ with equal probabilities.}
          \end{center}
        \end{figure}

To get the position of the critical temperature for our example
$q=3,4$ we use the basic knowledge about spontaneous magnetization.
The dependence of the spontaneous magnetization on temperature
provides an opportunity to single out the critical point. Iteration
process and evaluation of the magnetization under the zero magnetic
field, as it was described above, give us dependence of the
spontaneous magnetization on $\beta J$ (Fig. (\ref{crit})). The
critical temperature lies between the critical temperatures
corresponding to the same models on Bethe lattices with constant
$q=3$ and $q=4$ coordination numbers.

The same analysis may be done for the quenched topological disorder
with $q=2,3$. Our study of the spontaneous magnetization shows that
the critical point exists and the critical temperature differs from
zero even when the probability of $q=2$ is greater than the
probability of $q=3$ ($P(3)=0.01$, $2<\beta_{c} J<3$). It means that
the little possibility of branching in one dimensional Ising model
(rare branches) lead to the critical behavior with non zero critical
temperature in the thermodynamical limit.

 As it is easy to see, we mentioned above some propositions which
are very general (statements about attractor and other ) and they
need to be studied with mathematical rigor. However, this statements
are useful and numerical research gives an evidence of their
correctness.
\section{Conclusion}\label{5}
Summarizing written above, we developed an approach to deal with
topological disorder for the spin models on hierarchical lattices.
Despite its demonstration only on the simple Bethe lattice, a
generalization on more complicated models is straightforward. A wide
spectra of problems solving within approach of hierarchical models
and having well physical description within it may be revisited now
in presence of fluctuating coordination number of lattice. A
complicated disorder may lead to the nontrivial solutions of the
recursive relations Eq. (\ref{mapp}). The distributions of the
Yang-Lee and Fisher zeros in presence of disorder is also an
interesting problem. Speaking about social dynamics we see that the
hierarchical models take a possibility to simplify the problem
within Ising paradigm without significant loss of verity and
developed approach makes it possible to get the major interesting
quantities.
\section{Acknowledges}
 Authors are grateful to E. Mamasakhlisov and A. Allahverdyan for discussions
 . K.S. would like to thank K. Mazmanian for useful remarks.
This work was supported by ANSEF grant PS-condmatth-521.

\end{document}